# Semimetallic transport properties of epitaxially stabilized perovskite CaIrO$_3$ films

Daigorou Hirai[1*], Jobu Matsuno[2], and Hidenori Takagi[1,3]

[1]*Department of Physics, University of Tokyo, Hongo 7-3-1, Bunkyo-ku, Tokyo 113-0033, Japan*

[2] *RIKEN Center for Emergent Matter Science (CEMS), Wako, Saitama 351-0198, Japan*

[3]*Max-Plank-Institute for solid state research, Heisenbergstrasse 1, Stuttgart 70569, Germany*

E-mail: dhirai@qmat.phys.s.u-tokyo.ac.jp

We report on the synthesis and transport properties of perovskite (Pv) CaIrO$_3$ thin films. The Pv phase of CaIrO$_3$ was stabilized by epitaxial growth on SrTiO$_3$, (LaAlO$_3$)$_{0.3}$(Sr$_2$AlTaO$_6$)$_{0.7}$, and LaAlO$_3$ substrates with strong tensile, weak tensile, and compressive strains, respectively. The resistivity of these films showed a poorly metallic behavior. The Hall resistivity exhibited a sign change as a function of temperature and a nonlinear magnetic-field dependence, which clearly indicated the coexistence of electrons and holes and hence supported that Pv CaIrO$_3$ films are semimetallic. The observed robustness of the semimetallic ground state against tensile and compressive strains is consistent with the presence of symmetry-protected Dirac points (nodes) around the Fermi level that prohibits the system from becoming a band insulator.

A surge of interest in 5$d$ transition-metal oxides has been initiated by the discovery of a spin-orbit coupling (SOC)-induced Mott state in $Sr_2IrO_4$ [1,2], where the SOC, the Coulomb interaction $U$, and the crystal-field splitting $\Delta$ are comparable in energy, and therefore a novel interplay of these electronic parameters may be anticipated. $Sr_2IrO_4$ is in the two-dimensional limit ($n = 1$) of the Ruddlesden-Popper series $Sr_{n+1}Ir_nO_{3n+1}$ ($n$: integer). The ground state of the series changes from an SOC-induced Mott insulator to a metal with increasing $n$ [3]. Early studies considered the $n = \infty$ end member, perovskite (Pv) $SrIrO_3$, to be an ordinary correlated metal [3]. Recent theoretical and experimental works, however, revealed that the strong SOC and the crystal symmetry of Pv $SrIrO_3$ lead to an unexpected semimetallic state in close proximity to a SOC-induced Mott insulator [4]. Attempts to manipulate the semimetallic state of $SrIrO_3$ have been made by varying the thickness [5], epitaxial stain [6], and dimensionality using a superlattice structure [7] and the chemical pressure using cation substitution [8]. It has been suggested theoretically that a topological state may be realized in a superlattice structure [9,10].

At ambient pressure, $CaIrO_3$ adopts a post-Pv structure [11], known as the high-pressure phase of the Pv structure, due to the mismatch between the $Ca^{2+}$ ion and the $IrO_6$ octahedra. The Pv phase of $CaIrO_3$ could be marginally stabilized with a wet chemical method, but only as a powder sample [12]. The growth of Pv $CaIrO_3$ [pseudo-cubic lattice parameter ($a_{pc}$) ~ 3.855 Å] thin films was reported on $GdScO_3$(110) ($a_{pc}$ ~ 3.96 Å) and $SrTiO_3$(001) [cubic lattice parameter ($a$) = 3.905 Å] substrates, where a tensile epitaxial strain was exerted [8]. It has been demonstrated that the presence of symmetry-protected Dirac points (nodes) near the Fermi level inherently produce a semimetallic state in Pv $SrIrO_3$ [4]. Similarly to Pv $SrIrO_3$, Pv $CaIrO_3$ is also anticipated to be a semimetal, owing to the protection of the Dirac points by symmetry. Because of the large lattice distortion due to the smaller ionic radius of $Ca^{2+}$ than that of $Sr^{2+}$, the correlation effect could be enhanced in Pv

CaIrO$_3$, which suggests that Pv CaIrO$_3$ is in the closer vicinity to a Mott insulator compared with Pv SrIrO$_3$.

In this study, we investigated the transport properties of single crystalline Pv CaIrO$_3$ thin films on SrTiO$_3$ (STO), (LaAlO$_3$)$_{0.3}$(Sr$_2$AlTaO$_6$)$_{0.7}$ (LSAT) and LaAlO$_3$ (LAO) substrates with strong tensile, weak tensile, and compressive strain respectively. Perovskite CaIrO$_3$ was successfully stabilized epitaxially on all three substrates. The transport properties of Pv CaIrO$_3$ were found to be consistent with the presence of a semimetallic ground state. An observed sign change of the Hall coefficient as a function of temperature can be reasonably explained by assuming that (1) CaIrO$_3$ is a semimetal, where electron bands associated with the Dirac points (nodes) and hole bands coexist at the Fermi energy, and (2) holes are doped into the system owing to impurities/defects. The semimetallic character of Pv CaIrO$_3$ was found to be rather robust against epitaxial strains, which may suggest that the presence of Dirac points protected by lattice symmetry is the origin of the semimetallic ground state.

Thin films of CaIrO$_3$ with 10 to 30 nm thickness were epitaxially grown by pulsed laser deposition on (001)-oriented SrTiO$_3$ (STO) ($a$ = 3.905 Å), (LaAlO$_3$)$_{0.3}$(Sr$_2$AlTaO$_6$)$_{0.7}$ (LSAT) ($a$ = 3.868 Å), and LaAlO$_3$ (LAO) ($a$ = 3.793 Å) substrates. A KrF excimer laser ($\lambda$ = 248 nm) at 10 Hz with a fluence of ~ 1.5 J/cm$^2$ was used for the deposition. Polycrystalline pellets of post-Pv CaIrO$_3$, which were used as the target for the deposition, were prepared from CaCO$_3$ and IrO$_2$ by a conventional solid-state reaction. The substrate temperature and the oxygen partial pressure during the growth were 730°C and 25 Pa, respectively. The quality and the lattice parameters of the films were characterized by x-ray diffraction (XRD) measurements with Cu $K\alpha$ radiation using a diffractometer (Rigaku SmartLab) with a Ge(002) monochromator. The thickness of the films was estimated from the thickness fringes of the CaIrO$_3$(001) peak in x-ray $\theta$-$2\theta$ measurements and x-ray reflection measurements. The magnetic-field and temperature dependence of the resistivity and Hall coefficient were measured using a physical property measurement system (PPMS, Quantum Design).

The observation of well-defined pseudo-cubic 00$l$ ($l$ = 1 - 4) peaks in the $\theta$-$2\theta$ XRD patterns, shown in Fig. 1(a), indicated the growth of single crystalline films of (001)-oriented Pv CaIrO$_3$. No signature of impurity phases, such as post-Pv CaIrO$_3$, Ir, and IrO$_2$, was observed. The full width at half maximum of the 002 peak in the rocking curve was around 0.08°, indicative of the high crystallinity of the films. The pseudo-cubic lattice parameter of Pv CaIrO$_3$, estimated from the orthorhombic lattice parameters ($\sim\sqrt{2}a_{pc} \times \sqrt{2}a_{pc} \times 2a_{pc}$) of the bulk material, was ~ 3.858 Å [12]. The lattice mismatches between bulk CaIrO$_3$ and the substrates are listed in Table I. Strong tensile, weak tensile and compressive strains are expected to be exerted on the CaIrO$_3$ thin films on the STO, LSAT, and LAO substrates, respectively. In Fig. 1(b), the $\theta$-$2\theta$ XRD scans display a shift of the 001 peak to a higher angle with increasing lattice parameters of substrates from LAO to STO. This is consistent with the contraction of the out-of-plane lattice parameters associated with the in-plane tensile strain.

In order to evaluate the in-plane lattice constants of the films, reciprocal space mapping was conducted, as shown in Figs. 1(c) and 1(d), which supported the epitaxial growth of the thin Pv films. All CaIrO$_3$ films thinner than 10 nm were coherently strained to their substrates, while films thicker than 30 nm were not fully strained when the lattice mismatch is relatively large (LAO and STO). The strain-induced biaxial tetragonal distortions for the 30-nm films were quantitatively characterized by the ratio $c/a$ of the in-plane and out-of-plane lattice constants ($a$ and $c$) obtained from the $\theta$-$2\theta$ XRD scans and reciprocal space maps (Table I). Strong tensile, weak tensile, and compressive strains with *c*/*a* equal to 0.988, 0.998, and 1.007 were confirmed for the CaIrO$_3$ films on the STO, LSAT, and LAO substrates, respectively. It must be noted that a pseudo-cubic notation is used for all the CaIrO$_3$ films in this study. TEM studies for SrIrO$_3$ thin films on STO(001) substrates have revealed that the $\sqrt{2}a_{pc} \times \sqrt{2}a_{pc} \times 2a_{pc}$ orthorhombic structure is stabilized [13]. A similar orthorhombic distortion due to the rotation and tilting of the IrO$_6$ octahedra may be expected for Pv CaIrO$_3$.

The results of the transport measurements of the examined films were consistent with the semimetallic character of the ground state of CaIrO$_3$. Figure 2 shows the temperature dependence of the resistivity $\rho(T)$ for 30-nm-thickness CaIrO$_3$ films on various substrates. At 300 K, the resistivity was as large as 1 mΩcm, which is comparable to that reported for CaIrO$_3$ thin films on GdScO$_3$ [8] and one order of magnitude smaller than that reported for bulk polycrystalline CaIrO$_3$ [12]. All the films exhibited a poorly metallic $\rho(T)$ ($d\rho/dT \sim 0$) only at low *T*, followed by a broad shoulder and then semiconducting $\rho(T)$ ($d\rho/dT < 0$) at high temperatures. This is the typical behavior of a semimetal with a small number of electrons and holes: At low temperatures, the system behaves as a degenerate metal with small electron and hole Fermi surfaces. In contrast, $\rho(T)$ is dominated by the number of electrons and holes thermally excited at high temperatures.

Clear evidence for the coexistence of electrons and holes in CaIrO$_3$ was provided by the sign change of the Hall coefficient $R_H(T)$ [see Fig. 3(a)]. The sign change of $R_H$ as a function of temperature implies a very delicate balance between the electron and hole contributions. In addition, the strong non-linearity in the field dependence of the Hall resistivity $\rho_{xy}(B)$ at low temperatures, shown in Fig. 3(b), demonstrates a subtle balance in the competition between the electron and hole contributions to the transport.

The ground state of $SrIrO_3$ is very close to that of a band insulator due to the interplay of the strong SOC and the orthorhombic lattice distortion [4]. Detailed band calculations with SOC indicated a small overlap between electron bands with a Dirac dispersion and hole bands with a normal dispersion [4]. The light Dirac electrons and the heavy holes are expected to contribute to the charge transport. Since zero-gap Dirac bands are protected by the lattice symmetry, the system cannot be a band insulator and a robust semimetallic ground state is generated.

When both electron and hole contributions coexist, the Hall coefficient in the low-field limit can be represented as $R_H = 1/e \cdot (n_h \mu_h^2 - n_e \mu_e^2)/(n_e \mu_e + n_h \mu_h)^2$, where $n_e$ ($n_h$) and $\mu_e$ ($\mu_h$) are the concentration and the mobility of electrons (holes), correspondingly (see, for example, [14]). In the case of a fully compensated semimetal ($n_e = n_h$), such as Pv $SrIrO_3$, the sign of $R_H$ is determined by the relative motilities of the two types of carriers. The higher mobility of the Dirac electrons compared with that of the heavy holes ($\mu_e > \mu_h$) should yield a negative $R_H$ ($n_h \mu_h^2 - n_e \mu_e^2 < 0$) at the $T = 0$ limit. At high temperatures, because of the high density of states of the hole band, the dominant thermal excitation should be from the hole band to the conduction band associated with the Dirac point. Therefore, at any temperature, the Hall coefficient should have a negative value owing to the higher mobility of the Dirac electrons, which is consistent with the reported $R_H(T)$ in Pv $SrIrO_3$ [6].

We might anticipate essentially the same electronic structure, and therefore transport behavior, as in Pv $SrIrO_3$. In contrast to the case of Pv $SrIrO_3$, however, the $R_H(T)$ of Pv $CaIrO_3$ exhibited a sign change from negative to positive with decreasing temperature [Fig.

3(a)]. This does not necessarily mean that the electronic structure of Pv CaIrO$_3$ is different from that of Pv SrIrO$_3$. If we assume a sizable amount of hole doping due to some impurities/defects in Pv CaIrO$_3$, the contrasting behavior of $R_H(T)$ in CaIrO$_3$ can be described by the same electronic structure as that proposed for SrIrO$_3$.

With hole doping, the concentration of heavy holes increases while that of Dirac-like electrons is suppressed at low temperatures. In that case ($n_h > n_e$), despite the higher mobility of the Dirac-like electrons ($\mu_e > \mu_h$), the contribution of the heavy holes to $R_H$ could marginally exceed that of the electrons ($n_h\mu_h^2 >\sim n_e\mu_e^2$), which explains the positive $R_H$ in CaIrO$_3$ at low temperatures. Indeed, the pronounced non-linear magnetic-field dependence of the Hall resistivity $\rho_{xy}(B)$ [Fig. 3(b)] at low temperatures, which is absent in Pv SrIrO$_3$ [6], implies the presence of hidden but competing electron contributions at low temperatures and a stronger magnetic-field dependence of $\mu_h$ than that of $\mu_e$. This corroborates the hypothesis of a marginal competition between the hole and electron contributions in Pv CaIrO$_3$, although the origin of the stronger field dependence of $\mu_h$ is not trivial. At high temperatures, almost equal numbers of Dirac-like electrons and heavy holes are thermally excited and the number of electrons and holes tends to balance ($n_e \sim n_h$). Therefore, the high mobility of the Dirac-like electrons should yield a negative $R_H$ ($n_h\mu_h^2 - n_e\mu_e^2 < 0$), as observed in the experiment.

$\rho(T)$ and $R_H(T)$ did not show any evident dependence on the type of substrate, as shown in Fig. 3, indicating that the semimetallic state discussed above is quite robust against epitaxial strain; epitaxial strain could modify the band, for example, through Ir-O-Ir bond angle. This observation appears to support that the semimetallic ground state is inherently created by the presence of zero-gap Dirac bands protected by symmetry. Some change in $R_H(T)$ and $\rho_{xy}(B)$ can be observed at low temperatures, but it does not exhibit a systematic dependence on the lattice constant of the substrate, which very likely suggests that extrinsic factors, such as the amount of the dopant, are responsible for the change.

In conclusion, we stabilized single-crystalline Pv $CaIrO_3$ thin films on (001)-oriented STO, LSAT, and LAO substrates with strong tensile, weak tensile, and compressive strains. Transport measurements indicated the coexistence of electrons and holes, consistent with the semimetallic character of the ground state proposed theoretically. We argue that $CaIrO_3$ is a hole-doped semimetal with Dirac electrons and heavy holes, which provides a reasonable description of the observed behavior of the resistivity and the Hall effect. The transport properties were found to be rather insensitive to epitaxial strains, which may suggest that the semimetallic ground state is robust owing to the presence of the protected Dirac points (nodes). Our findings give an insight into the unusual semimetallic ground state produced by the interplay of SOC and lattice distortion, where Dirac physics plays a role.

**Acknowledgements**

The authors thank Hae-Young Kee for the insightful discussion. This work was supported by a Grant-in-Aid for Scientific Research (grant number 24224010 and 25103724) from MEXT, Japan.

**References**

[1] B. J. Kim, H. Jin, S. J. Moon, J.-Y. Kim, B.-G. Park, C. S. Leem, J. Yu, T. W. Noh, C. Kim, S.-J. Oh, J.-H. Park, V. Durairaj, G. Cao, and E. Rotenberg, Phys. Rev. Lett. **101**, 076402 (2008).

[2] B. J. Kim, H. Ohsumi, T. Komesu, S. Sakai, T. Morita, H. Takagi, and T. Arima, Science **323**, 1329 (2009).

[3] S. J. Moon, H. Jin, K. W. Kim, W. S. Choi, Y. S. Lee, J. Yu, G. Cao, A. Sumi, H. Funakubo, C. Bernhard, and T. W. Noh, Phys. Rev. Lett. **101**, 226402 (2008).

[4] M. A. Zeb and H.-Y. Kee, Phys. Rev. B **86**, 085149 (2012).

[5] F.-X. Wu, J. Zhou, L. Y. Zhang, Y. B. Chen, S.-T. Zhang, Z.-B. Gu, S.-H. Yao, and Y.-F. Chen, J. Phys.: Condens. Matter **25**, 125604 (2013).

[6] J. Liu, J.-H. Chu, C. R. Serrao, D. Yi, J. Koralek, C. Nelson, C. Frontera, D. Kriegner, L. Horak, E. Arenholz, J. Orenstein, A. Vishwanath, X. Marti, and R. Ramesh, arXiv:1305.1732 [cond-Mat] (2013).

[7] J. Matsuno, K. Ihara, S. Yamamura, H. Wadati, K. Ishii, V. V. Shankar, H.-Y. Kee, and H. Takagi, arXiv:1401.1066 [cond-Mat] (2014).

[8] S. Y. Jang, H. Kim, S. J. Moon, W. S. Choi, B. C. Jeon, J. Yu, and T. W. Noh, J. Phys.: Condens. Matter **22**, 485602 (2010).

[9] D. Xiao, W. Zhu, Y. Ran, N. Nagaosa, and S. Okamoto, Nat. Commun. **2**, 596 (2011).

[10] J.-M. Carter, V. V. Shankar, M. A. Zeb, and H.-Y. Kee, Phys. Rev. B **85**, 115105 (2012).

[11] F. Rodi and D. Babel, Z. Anorg. Allg. Chem. **336**, 17 (1965).

[12] J.-G. Cheng, J.-S. Zhou, J. B. Goodenough, Y. Sui, Y. Ren, and M. R. Suchomel, Phys. Rev. B **83**, 064401 (2011).

[13] L. Zhang, H.-Y. Wu, J. Zhou, F.-X. Wu, Y. B. Chen, S.-H. Yao, S.-T. Zhang, and Y.-F. Chen, Appl. Surf. Sci. **280**, 282 (2013).

[14] *Semiconductors*, edited by R. A. Smith (Cambridge University Press, Cambridge, England, 1978).

**Figure Captions**

**Fig. 1.** (color online). (a) X-ray diffraction $\theta$-$2\theta$ scans of the 10-nm CaIrO$_3$ films on (001)-oriented LaAlO$_3$ (LAO), (LaAlO$_3$)$_{0.3}$(Sr$_2$AlTaO$_6$)$_{0.7}$ (LSAT), and SrTiO$_3$ (STO) substrates. Asterisks (*) represent peaks from the substrates. (b) Magnified area [dotted square in (a)] to show the change in lattice parameters for different substrates. Solid lines represent dynamical theory diffraction simulations around the pseudo-cubic (001) peaks. Reciprocal-space maps around the pseudo-cubic (103) reflections for (c) 10 nm and (d) 30 nm CaIrO$_3$ films, where $Q_{100}$ and $Q_{001}$ are in-plane and out-of-plane lattices, respectively.

**Fig. 2.** (color online). Temperature dependence of the resistivity for 30-nm CaIrO$_3$ films on LAO, LSAT, and STO substrates, respectively.

**Fig. 3.** (color online). (a) Temperature dependence of the Hall coefficient $R_H(T)$ and (b) magnetic-field dependence of Hall resistivity $\rho_{xy}(B)$ for CaIrO$_3$ films on (001)-oriented LAO, LSAT, and STO substrates, respectively. Solid lines in (a) show the linear fit for $\rho_{xy}(B)$ in the low-field limit.

Table I. In-plane lattice mismatch [($a_{substrate}$-$a_{bulk}$)/$a_{substrate}$], in-plane lattice parameter ($a$), out-of-plane lattice parameter ($c$), biaxial distortion ($c/a$), and unit cell volume for $Z = 1$ ($V$) for CaIrO$_3$ films (30 nm) on LAO, LSAT, and STO substrates, respectively.

| substrate | mismatch | $a$ (Å) | $c$ (Å) | $c/a$ | $V$(Å$^3$) |
|---|---|---|---|---|---|
| *Bulk | - | 3.858 | - | - | 57.43 |
| on LAO | -1.78% | 3.843(4) | 3.872(4) | 1.007 | 57.20 |
| on LSAT | +0.32% | 3.871(1) | 3.865(5) | 0.998 | 57.90 |
| on STO | +1.18% | 3.889(5) | 3.841(2) | 0.988 | 58.10 |

*The bulk lattice parameters were converted from the orthorhombic ones in ref [12].

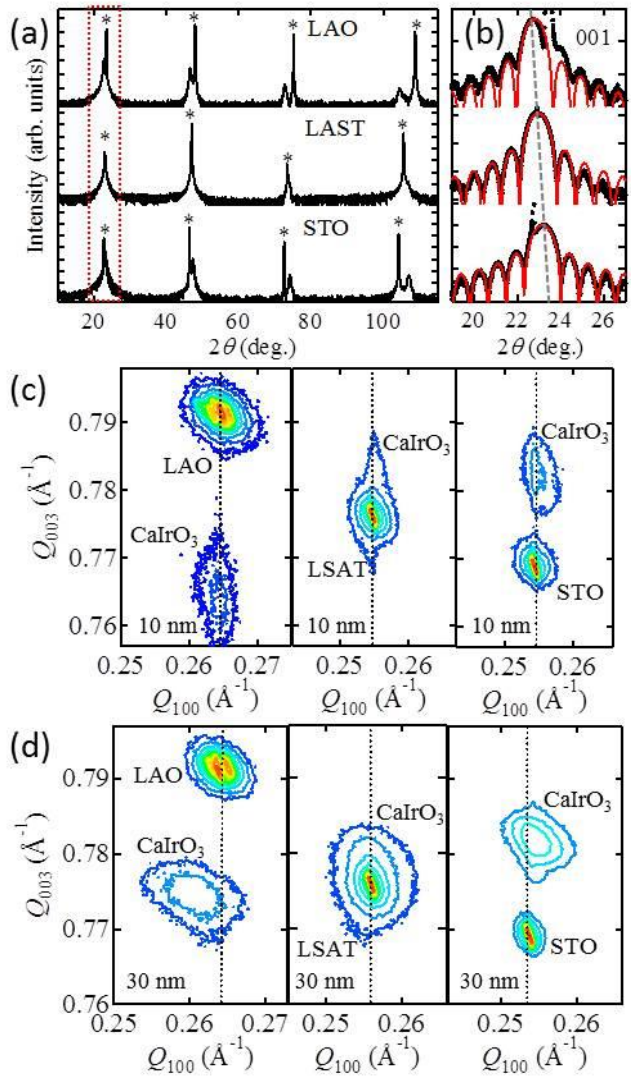

Fig. 1.

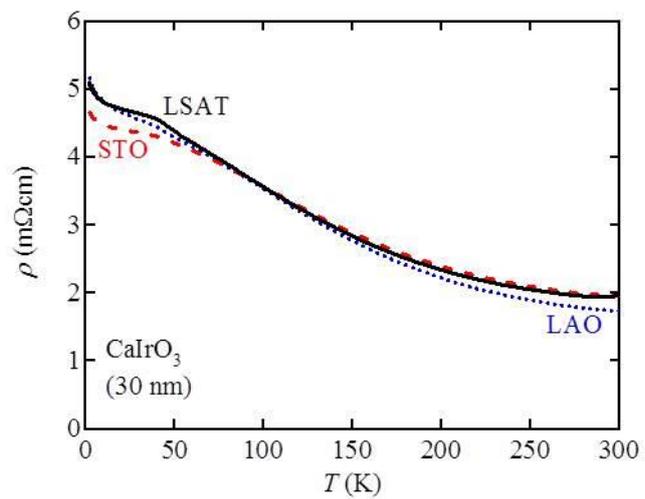

Fig. 2.

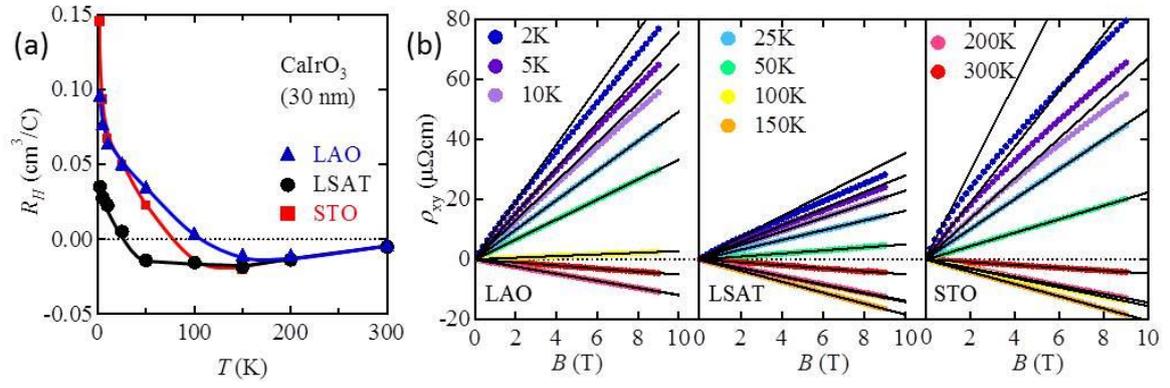

Fig. 3.